\catcode`\@=11
\def\citen#1{\if@filesw \immediate\write \@auxout {\string\citation{#1}}\fi%
\@tempcntb\m@ne \let\@h@ld\relax \def\@citea{}%
\@for \@citeb:=#1\do {\@ifundefined {b@\@citeb}%
    {\@h@ld\@citea\@tempcntb\m@ne{\bf ?}%
    \@warning {Citation `\@citeb ' on page \thepage \space undefined}}%
    {\@tempcnta\@tempcntb \advance\@tempcnta\@ne
    \setbox\z@\hbox\bgroup\ifcat0\csname b@\@citeb \endcsname \relax
    \egroup \@tempcntb\number\csname b@\@citeb \endcsname \relax
    \else \egroup \@tempcntb\m@ne \fi \ifnum\@tempcnta=\@tempcntb
    \ifx\@h@ld\relax \edef \@h@ld{\@citea\csname b@\@citeb\endcsname}%
    \else \edef\@h@ld{\hbox{--}\penalty\@highpenalty
    \csname b@\@citeb\endcsname}\fi
    \else \@h@ld\@citea\csname b@\@citeb \endcsname \let\@h@ld\relax \fi}%
\def\@citea{,\penalty\@highpenalty\hskip.13em plus.13em minus.13em}}\@h@ld}
\def\@citex[#1]#2{\@cite{\citen{#2}}{#1}}%
\def\@cite#1#2{\leavevmode\unskip\ifnum\lastpenalty=\z@\penalty\@highpenalty\fi%
  \ [{\multiply\@highpenalty 3 #1%
  \if@tempswa,\penalty\@highpenalty\ #2\fi}]}   %
\makeatother 

\catcode`\@=12

\def\A             {{\rm A}}
\def\alg           {algebra}
\def\auto          {automorphism}
\def\bc            {boundary condition}
\def\be            {\begin{equation}}
\def\bearl         {\begin{array}{l}}
\def\bearll        {\begin{array}{ll}}
\def\cala          {{\wfont A}}

\def\calap         {\bar{\wfont A}}
\def\calc          {{\cal C}}
\def\calh          {{\cal H}}
\def\calm          {{\cal M}}

\def\calq          {{\cal Q}}
\def\cals          {{\cal S}}
\def\calw          {{\cal W}}
\def\cblock        {chiral block}
\def\cft           {conformal field theory}
\def\cfts          {conformal field theories}
\def\chii          {\raisebox{.15em}{$\chi$}}

\def\class         {classification}
\def\complex       {{\dl C}}

\def\dl            {\mathbb }
\def\DPN           {\widehat{\rm N}}
\newcommand\DpNl[3]{\widehat{\rm N}_{#1,#2,#3}^{}}
\newcommand\DpNL[3]{\widehat{\rm N}_{#1,#2,#3}}
\def\dsty          {\displaystyle}
\def\ee            {\end{equation}}
\def\eE            {{\rm e}}
\def\eear          {\end{array}}
\def\eq            {\,{=}\,}
\newcommand\erf[1] {(\ref{#1})}
\newcommand\Erf[2] {(\ref{#1#2})}
\def\findim        {finite-dimensional}
\newcommand\Frac[2]{\mbox{\large$\frac{#1}{#2}$}}

\def\furu          {fusion rule}
\def\futnote#1     {\footnote{~#1}\ }

\def\gg            {\omega}
\def\Gs            {{G^*}}
\newcommand\hsp[1] {\mbox{\hspace{#1 em}}}
\def\hy            {$\mbox{-\hspace{-.66 mm}-}$}

\def\ii            {{\rm i}}
\def\iN            {\,{\in}\,}

\def\irr           {irreducible }
\def\irrep         {irreducible representation}
\def\J             {{\rm J}}
\def\kzc           {Knizh\-nik\hy Za\-mo\-lod\-chi\-kov connection}
\long\def\labl#1   {\label{#1}\ee}
\long\def\Labl#1#2 {\label{#1#2}\ee}
\def\lambdab       {{\bar\lambda}}
\def\lambdaB       {{[\lambdab]}}

\def\lie           {Lie algebra}
\def\llb           {\mbox{\LARGE(}}
\def\lrb           {\mbox{\LARGE)}}
\def\mub           {{\bar\mu}}
\def\muB           {{[\mub]}}

\def\NN            {{N_{(1)}}}
\newcommand\nxt[1] {\\\raisebox{.12em}{\rule{.35em}{.35em}}\hsp{.6}#1}

\def\onedim        {one-dimen\-sional}
\def\q             {quantum }
\def\rep           {representation}
\def\resp          {respectively}
\def\rhob          {\bar\rho}
\def\rhoB          {{[\rhob]}}
\def\rhoBd         {{[\rhob_3^{}]}}
\def\rhoBdP        {{[\rhob_3^+]}}
\def\rhoBe         {{[\rhob_1^{}]}}
\def\rhoBv         {{[\rhob_4^{}]}}
\def\rhoBz         {{[\rhob_2^{}]}}
\def\rhoBzP        {{[\rhob_2^+]}}
\def\rhod          {{\rho_3}}

\def\rhs           {right hand side}
\newcommand\sect[1]{\addtocounter{section}1%
                   \mbox{$ $}\\[-.5em]{\bf\thesection.\,\ #1}\\[.22em]}
\def\sigmab        {{\bar\sigma}}
\def\sigmaB        {{[\sigmab]}}

\def\sigmaBP       {{[\sigmab^+]}}
\def\sigmaBd       {{[\sigmab_3^{}]}}
\def\sigmaBdP      {{[\sigmab_3^+]}}
\def\sigmaBe       {{[\sigmab_1^{}]}}
\def\sigmaBeP      {{[\sigmab_1^+]}}
\def\sigmaBz       {{[\sigmab_2^{}]}}
\def\SJ            {S^\J}
\def\sltwo         {\mbox{$\mathfrak{sl}(2)$}}
\def\ssty          {\scriptstyle}
\newcommand\sumbo[1]{\sum_{\ssty\bar #1 \atop Q(#1)=0}}
\def\tA            {\bar{\rm A}}
\def\tC            {\Tilde C}
\def\Tilde         {\tilde}
\newcommand\tM[3]  {\bar{\rm M}^{#1,#2}_{\ \ \ \ \ \ \ \ \ \ \ #3}}
\def\TM            {\bar{\rm M}}
\newcommand\tN[3]  {\Tilde{\rm N}_{#1,#2}^{\ \ \ \ \ \ \ \ \ #3}}
\def\TN            {\Tilde{\rm N}}
\newcommand\tNl[3] {\Tilde{\rm N}_{#1,#2,#3}^{}}
\def\tS            {\Tilde S}

\def\vac           {\Omega}
\def\vacb          {{\bar\Omega}}

\def\wfont         {\mathfrak }
\def\wrt           {with respect to }

\def\wzwm          {WZW model}

\def\wzwts         {WZW theories}
\def\X             {{\cal X}}
\def\zet           {{\dl Z}}

\documentclass[12pt]{article} \usepackage{amssymb,amsfonts}
\begin{document}

\begin{flushright}  {~} \\[-1cm] {\sf hep-th/9811211} \\[1mm]
{\sf CERN-TH/98-371} \\[1 mm] {\sf November 1998} \end{flushright}
 
\begin{center} \vskip 22mm
{\Large\bf ORBIFOLD ANALYSIS OF}\\[4mm]
{\Large\bf BROKEN BULK SYMMETRIES}\\[22mm]
{\large J\"urgen Fuchs} \ and \ 
{\large Christoph Schweigert} \\[3mm] CERN \\[.6mm] CH -- 1211~~Gen\`eve 23
\end{center}
\vskip 22mm
\begin{quote}{\bf Abstract}\\[1mm]
In two-dimensional conformal field theory, we analyze conformally invariant 
boundary conditions which break part of the bulk symmetries. 
When the subalgebra that is preserved by the boundary conditions is the fixed 
algebra under the action of a finite group $G$, orbifold techniques can be 
used to determine the structure of the space of such boundary conditions.
We present explicit results for the case when $G$ is abelian.
In particular, we construct a classifying algebra which 
controls these symmetry breaking boundary conditions in the same way in which
the fusion algebra governs the boundary conditions that preserve the full
bulk symmetry.
\end{quote}
\newpage
 

\sect{Boundary conditions and consistent chiral algebras}
Conformal field theories on surfaces with boundaries have recently
attracted renewed interest. It was known for quite some time that such theories
play an important role in the analysis of condensed matter
systems, like e.g.\ in the Kondo effect, as well as in critical percolation.
An additional motivation to study this problem
was given by the discovery \cite{polc3} that 
string perturbation theory in the background of certain solitonic
solutions that describe black D-branes can be described in terms of 
open strings with non-trivial boundary conditions.
Thus by studying the space of conformally invariant boundary conditions for
the \cfts\ that constitute string vacua
one can obtain information about the possible solitonic sectors 
of string theory. Also, it is this space of \bc s (and the space of all 
possible crosscaps \cite{prss2,prad}) on which the problem of tadpole
cancellation should be considered.

Ideally, one would therefore like to study the space of {\em all\/} conformally 
invariant boundary conditions in any given \cft\ model. Unfortunately, except 
for particularly simple
models, this space does not seem to be tractable at the moment.
To handle this \class\ problem in the general case, one should 
start by grouping the various \bc s in a coarse manner into subspaces, and
then try a finer \class\ for each of these subspaces.
As a reasonable approach to the first step, we propose to characterize these
classes of boundary conditions by 
associating to each boundary condition the {\em subalgebra\/}
$\calap$ {\em of the chiral algebra\/} $\cala$ of the theory that is preserved
by the boundary condition. The requirement that the boundary condition has to
be conformally invariant means that $\calap$ must contain the Virasoro 
subalgebra of $\cala$. Furthermore, the subalgebra $\calap$ has to be a consistent
chiral algebra in the sense that the corresponding \cblock s, as vector
bundles over the moduli space of complex curves and insertion points, come with
a \kzc\ and obey suitable factorization rules. 

The special case when the boundary conditions preserve the full chiral 
algebra $\cala$ has received attention already long ago. As first argued 
by Cardy \cite{card9}, in this case the consistent
boundary conditions are in one-to-one correspondence with the (generalized)
quantum dimensions of the theory, i.e.\ with the \onedim\ \irrep s of the
fusion algebra. Typically a chiral algebra $\cala$ will, however, possess 
very many, if not infinitely many, consistent subalgebras $\calap$. 
The first step towards a classification of all boundary
conditions would be to classify all these subalgebras. This problem clearly 
depends largely on the specific bulk \cft\ under consideration, and we will
not have to say much about it in this letter.

The goal of this letter is, rather, to classify all those boundary conditions 
that preserve some prescribed subalgebra $\calap$. As long as $\calap$ is
completely arbitrary, at present this problem is still too general to be 
tractable. We will therefore restrict our attention to a particular subclass
of consistent subalgebras. Namely, we require
that $\calap$ is the {\em fixed algebra\/} of some group $G$ of automorphisms
of the chiral algebra $\cala$. In other words, $\calap\eq\cala^G$ is the 
chiral algebra of an orbifold of the theory that has chiral \alg\ $\cala$. 
The orbifold group $G$ need not necessarily be
finite, it can even be a \findim\ Lie group. Still, for the purpose of the 
present letter we specialize further to the case when $G$ is a finite
abelian group. This situation may seem rather special compared to the 
general problem sketched above, but it nevertheless covers a variety of 
examples of practical interest. Moreover, a number of physical insights can 
be gained, e.g.\ concerning the relation between \bc s that preserve sub\alg s
$\calap_1$ and $\calap_2$ of $\cala$ which are contained in each other.

\sect{Examples}
Let us present a number of concrete theories which realize the
situation described in the introduction. Our first example is the $c\eq1$ \cft\ 
of a free boson $X(z,\bar z)$ compactified on a circle. Its chiral algebra 
is generated by all polynomials in $\ii\partial X(z)$ as well as, in the 
rational case, certain normal-ordered exponentials $\exp(\ii k X(z))$, where 
the momentum $k$ lies on a lattice that depends on the compactification radius
$R$. The map $\gg{:}\; X\,{\mapsto}\,{-}X$ induces a symmetry of this chiral 
algebra. The corresponding orbifold theory is the well-known 
$\zet_2$-orbifold of the free boson.\,%
 \futnote{For $d$ uncompactified bosons, the group $\zet_2$ gets replaced 
by the Lie group O($d$).}
One can recover the original compactified free boson theory
by extending the chiral algebra of the orbifold theory by the field
$j\eq\ii\partial X$ which has \q dimension one and conformal weight one.
Correspondingly, we have two types of \bc s;
conditions of the first type preserve the whole chiral algebra 
$\cala$, while those  of the second type preserve only the subalgebra 
$\cala^{\zet_2}_{}$ that is fixed under $\gg$. This distinction is again 
well known: the first type are Neumann boundary conditions,
while the second are Dirichlet conditions, \resp\ the other way round 
(the two situations are exchanged by charge conjugation, i.e.\ T-duality). 

The free boson at rational radius squared provides yet another example. 
One can see that actually only D-branes sitting at suitable roots of unity 
preserve the full rational symmetry of the theory (\resp\ of its
$\zet_2$-orbifold). To obtain also D-branes at generic
locations, one has to break the bulk symmetry in the following manner. 
Instead of including all the exponentials $\exp(\ii k X)$ with $k$ in the 
relevant lattice, one restricts the allowed values of $k$ to a sublattice.
The so obtained sub\alg\ $\calap$ of $\cala$ is precisely the chiral \alg\
of a free boson theory whose compactification radius is an integral
multiple $MR$ of the original one; it can be described as the \alg\
$\cala^{\zet_M}_{}$ that is invariant under the $\zet_M$ group of \auto s
generated by the shift $X\,{\mapsto}\,X+2\pi/M\sqrt N$, where $N$ is the 
number of primary fields of the original theory.

Another example \cite{afos,fuSc9,bePz}
is the three-state Potts model which has a $\calw_3$-symmetry.
The boundary conditions which preserve the whole
$\calw_3$-symmetry are the so-called fixed and mixed \bc s. 
The $\calw_3$-algebra has an automorphism $\gg$ of order two that maps the 
spin-three current to minus itself; in the Potts model the fixed subalgebra 
\wrt $\gg$ is just the Virasoro algebra; the \bc s which preserve only the 
Virasoro algebra are the free boundary condition as well as the new
boundary condition discovered in \cite{afos}.
A similar situation arises for all Virasoro minimal models with central
charge $c=1-6/m(m+1)$ for $m\eq1$\,or\,$2\bmod 4$ and with modular invariant of
extension type \cite{fuSc9}. 

A different class of examples is provided by \cfts\ which are tensor products 
of identical subtheories. Then there are boundary conditions which preserve
only the subalgebra that is fixed under a cyclic group of permutations of the
subtheories. Such \bc s can be analyzed by combining our results with the 
methods developed in \cite{bohs,bant8}.

Finally we mention that when talking about \bc s one usually refers to the
situation where the torus partition function is the charge conjugation
modular invariant. 
T-duality, on the other hand, maps the \bc s for the true diagonal modular 
invariant that respect all bulk symmetries to those boundary conditions
for the charge conjugation modular invariant that are twisted by charge
conjugation. Applying the formalism developed in this paper to the orbifold by 
the $\zet_2$-symmetry that is furnished by charge conjugation therefore allows 
in particular to determine boundary conditions for the true diagonal modular 
invariant.

\sect{Simple current extensions -- a summary}
The chiral algebra $\cala$ can be decomposed 
into eigenspaces for the action of the finite abelian orbifold group
$G$. These eigenspaces are labelled by characters $\Psi$ of $G$, i.e.\ we have
  \be  \cala= \bigoplus_{\Psi\in\Gs} \cala_\Psi \,. \labl0
The fixed \alg\ $\calap$ can be identified with the eigenspace for the trivial 
character $\Psi_0\iN\Gs$, $\calap\eq\cala_{\Psi_0}$; the other eigenspaces are
modules of $\calap$. Inspection shows that all the examples presented 
above share another important feature: the spaces $\cala_\Psi$ appearing in 
\erf0 are even {\em irreducible\/} $\calap$-modules. In fact, we are
not aware of any abelian orbifold theory for which this property does not hold;
accordingly we will from now on assume that indeed all $\cala_\Psi$ are
irreducible.

This assumption implies in particular
that in the orbifold theory the fusion rules of the primary fields that
correspond to the modules
$\cala_\Psi$ are given by the character group $\Gs$ of $G$, which is again
a finite abelian group. In other words, all these primary fields of the
orbifold theory are {\em simple currents\/} \cite{scya,scya6}, and the original
chiral \alg\ $\cala$ can be recovered from $\calap$ as a so-called
integer spin simple current extension. This observation
enables us to use simple current technology to investigate the problem.
The rest of this section is devoted to a brief review of the properties of
simple current extensions as established in \cite{scya6,fusS6} that will be
needed in the sequel.

Let us consider the following situation in chiral\,%
 \futnote{At the chiral level, where one deals with \cft\ on a complex curve,
there is  no influence of boundaries at all \cite{fuSc6}. 
The chiral \cft\ structures considered here are thus logically 
independent of any boundary data; they have passed independent tests
\cite{fusS6,bant6} in the context of closed \cft.}
\cft. We start with some chiral \cft\ with chiral algebra $\calap$; the fusion
\alg\ of this theory has the character group $\Gs$ as a subgroup, whose
elements $\J$ correspond to integer spin simple currents. The theory obtained 
by extending $\calap$ by these simple currents is precisely the theory with 
chiral algebra $\cala$.  The simple currents $\J\iN\Gs$ act via the fusion 
product on the primary fields\,%
 \futnote{To be precise, on the corresponding generators $\phi_\lambdab$
of the fusion \alg.}
$\lambdab$ of the $\calap$-theory; this action organizes 
them into orbits $[\lambdab]$. 

We need the following additional data. To every primary field
$\lambdab$ we associate its stabilizer 
  \be  \cals_\lambdab := \{\J\iN\Gs\,|\, \J\,{\star}\,\lambdab\eq\lambdab \}
  \, . \ee
Every $\cals_\lambdab$ is a subgroup of $\Gs$, and it is one and the same 
subgroup for fields on the same $\Gs$-orbit. When $\J\iN\cals_\lambdab$, 
we say that $\lambdab$ is a {\em fixed point\/} of the simple current $\J$.
Further, to every simple current $\J\iN\Gs$
and to every field $\lambdab$ one associates the {\em monodromy charge\/}
  \be  Q_\J(\lambdab):= \Delta_\lambdab+ \Delta_\J - \Delta_{\J\star\lambdab}
  \bmod\zet \, . \labl4
Moreover, for every simple current $\J\iN\Gs$ we have a matrix $\SJ$ whose
entries $\SJ_{\lambdab,\mub}$ are non-vanishing only if both primaries
$\lambdab$ and $\mub$ of the $\calap$-theory are fixed points of $\J$, 
i.e.\ only if both $\J\lambdab\,{\equiv}\,\J\,{\star}\,\lambdab\eq\lambdab$ and 
$\J\mub\eq\mub$. For the identity element $1\iN\Gs$, $S^1\eq\bar S$ is the 
ordinary modular S-matrix of the $\calap$-theory.

The restriction of $\SJ$ to the fixed points of $\J$ is unitary, and
together with the restriction of the $T$-matrix it obeys
the usual relations of the modular
group; further, it satisfies the simple current relation
  \be  \SJ_{\J'\lambdab,\mub} = \eE^{2\pi\ii Q_{\J'}(\mub)} \SJ_{\lambdab,\mub}
  \labl2
for every simple current $\J'\iN\Gs$. As a matter of fact, in full
generality the relation \erf2 only holds up to a certain
two-cocycle on $\Gs$. This forces one to deal also with a subgroup
of the stabilizer on which the cocycle vanishes, the {\em untwisted 
stabilizer\/} \cite{fusS6,bant6,bant7}, rather than only
with the full stabilizer. In order to present our results without
much additional notation, for the purposes of this letter we 
will ignore this important complication. For a detailed description, 
with full account of the untwisted stabilizer, we refer to a
forthcoming publication \cite{fuSc+}.

There is evidence \cite{bant6} that the matrix $\SJ$ coincides with the matrix
that implements the modular transformation $\tau\,{\mapsto}\,{-}1/\tau$ on the 
one-point chiral blocks on the torus with insertion $\J$. When the
$\calap$-theory is a \wzwm\ or a coset model, then the matrix $\SJ$ is the
Kac\hy Peterson matrix of the relevant orbit \lie, see \cite{fusS3,furs}.
In the case of our present interest, for a large 
class of \cfts\ we can also use the result \cite{zhu3} that under certain
finiteness conditions one can associate to every descendant of the vacuum
a representation of the modular group; this result is relevant here because
in the extended theory with chiral algebra $\cala$ the simple currents in $\Gs$ 
become descendants of the vacuum.

Under the restriction that all untwisted stabilizers equal the full stabilizers,
the pertinent results of \cite{fusS6} can be summarized as follows.
\nxt 
The primary fields of the $\cala$-theory are (labelled by) pairs
$(\lambdaB{,}\psi_\lambda)$, where $\lambdaB$ is a
$\Gs$-orbit with vanishing monodromy charge, $Q_\J(\lambda)\eq0$ for all
$\J\iN\Gs$,\,%
 \futnote{Standard simple current relations imply that for every simple current
$\J$ of integral conformal weight the monodromy charges $Q_\J(\lambdab)$ 
defined by \erf4 are constant on $\Gs$-orbits.
As already mentioned, the same is true for the stabilizer subgroups. We 
therefore simplify notation by writing $Q(\lambda)$, $\psi_\lambda$, 
$\cals_\lambda$ etc.\ in place of $Q(\lambdab)$ etc.} 
and where $\psi_\lambda$ is a character of the stabilizer, 
$\psi_\lambda\iN\cals_\lambda^*$.
\nxt 
It follows in particular that an \irr module $\calh_{([\lambdab],\psi_\lambda)}$
of the $\cala$-theory decomposes into \irr $\calap$-modules 
$\bar\calh_\mub$ according to
  \be  \calh_{([\lambdab],\psi_\lambda)} = \bigoplus_{\J\in\Gs/\cals_\lambda}
  \bar\calh_{\J\lambdab} \,.  \labl1
(In the special case where $\lambdab\eq\vacb$ is the vacuum of the 
$\calap$-theory,
which has monodromy charge zero and is on a full $\Gs$-orbit, this is nothing 
but \erf0.) Notice that for non-trivial stabilizer one and the same
$\calap$-module $\bar\calh_\mub$ can appear in the decomposition of 
several distinct \irr $\cala$-modules.
\nxt 
The modular matrix $S$ of the $\cala$-theory is given by
  \be  S_{([\lambdab],\psi_\lambda) ([\mub],\psi_\mu)}
  = \Frac{|\Gs|}{|\cals_{\lambda}|\,|\cals_\mu|} 
  \sum_{\J\in\cals_\lambda\cap\cals_\mu}
  \psi_\lambda(\J)\, \psi_\mu(\J)^* \, \SJ_{\lambdab,\mub} \,.  \labl6

\sect{The classifying algebra}
By the requirement that $\calap$ is a consistent chiral algebra, the
chiral blocks of the orbifold theory satisfy the usual factorization rules.
This allows \cite{reSC,fuSc6} to analyze the factorization of bulk-bulk-boundary
correlators \cite{lewe3,prss3,sasT2} 
in the same manner as for \bc s which preserve all of $\cala$. This way 
one obtains \cite{fuSc6} the reflection coefficients for a bulk 
field in the presence of any conformally invariant boundary condition
from the \onedim\ \irrep s of a certain algebra, 
the {\em classifying algebra\/} $\calc(\calap)$. The structure constants of 
$\calc(\calap)$ can be expressed in terms of the operator product coefficients 
of the $\cala$-theory and of fusing matrices for the boundary blocks. Such 
fusing matrices exist because by assumption the chiral blocks of the 
$\calap$-theory possess a \kzc.

The classifying algebra $\calc(\cala)$ for those \bc s which
preserve all of $\cala$ is just the fusion \alg\ of the $\cala$-theory.
Accordingly a basis of $\calc(\cala)$ is given by the primary fields 
$(\lambdaB{,}\psi_\lambda)$ of the $\cala$-theory. 
On the other hand, in the presence of boundary conditions which preserve 
only a proper sub\alg\ $\calap$ of the $\cala$-symmetry, different submodules 
$\bar\calh_\mub$ in the decomposition \erf1 are reflected differently 
at the boundary. To take this behaviour into account, as 
a basis of the classifying algebra $\calc(\calap)$ we then take 
{\em individual\/} \irr $\calap$-{\em modules\/} 
rather than orbits of $\calap$-modules. Nevertheless
we also have to take the characters $\psi_\lambda$ into account, because
one and the same \irr $\calap$-module is reflected differently when it appears
in different $\cala$-modules $\calh_{\muB,\psi_\mu}$. In short, the basis 
elements of $\calc(\cala)$ must be labelled by pairs 
$(\lambdab{,}\psi_\lambda)$, where $\lambdab$ is an $\calap$-primary with 
vanishing monodromy charge and $\psi_\lambda$ is a character of the stabilizer 
$\cals_\lambda$.
The set of these fields is closely related to the set of primaries in the
untwisted sector of the orbifold theory based on $\calap$; but it is not exactly
the same, since we include multiplicities (encoded in the characters 
$\psi_\lambda$) for those fields in the untwisted sector that appear more than
once in the $\cala$-theory.

To obtain the structure constants of the classifying algebra, in principle one 
could now proceed as described in \cite{prss3,fuSc6} and work out the 
factorization of bulk-bulk-boundary correlators. 
Unfortunately, except for a few special cases the required values of
operator product coefficients and fusing matrices are not known.
However, we can circumvent this problem entirely by combining the information
about $\calc(\calap)$ and its basis given above with our knowledge about
simple current extensions.

This way we arrive at the following results (for details of the calculations, 
and also for a proper treatment of genuine untwisted stabilizers, see 
\cite{fuSc+}). Let us first present the structure constants $\TN$ of
$\calc(\calap)$ with only lower indices; we have
  \be \tNl{(\lambdab_1,\psi_1)}{(\lambdab_2,\psi_2)}{(\lambdab_3,\psi_3)}
  = \Frac{|\Gs|}{ |\cals_{\lambda_1}{\cdot}\,\cals_{\lambda_2}{\cdot}\,
  \cals_{\lambda_3}|}
  \,\DpNl{(\lambdab_1,\psi_1)}{(\lambdab_2,\psi_2)}{(\lambdab_3,\psi_3)}
  \,,  \ee
where various quantities are introduced as follows. By 
$\cals_{\lambda_1}{\cdot}\,\cals_{\lambda_2}{\cdot}\,\cals_{\lambda_3}$ 
we denote the subgroup of $\Gs$ that 
is generated by the three stabilizers $\cals_{\lambda_i}$. The quantity
$\DpNL{(\lambdab_1,\psi_1)}{(\lambdab_2,\psi_2)}{(\lambdab_3,\psi_3)}$ is the 
rank of a natural subsheaf of the bundle of \cblock s of the $\cala$-theory 
with insertions $([\lambdab_1]{,}\psi_1),\; ([\lambdab_2]{,}\psi_2)$ and 
$([\lambdab_3]{,}\psi_3)$. More explicitly, $\DPN$ is given by the Verlinde-like 
formula \cite{fuSc8}
  \be  \DpNl{(\lambdab_1,\psi_{\lambda_1})}{(\lambdab_2,\psi_{\lambda_2})}
  {(\lambdab_3,\psi_{\lambda_3})}
  = \!\sum_{\J_1\in\cals_{\lambdab_1}} \sum_{\J_2\in\cals_{\lambdab_2}}
  \sum_{\J_3\in\cals_{\lambdab_3}}
  \!\!\Frac{ \delta^{}_{\J_1\J_2\J_3,1} }\NN \,
  \llb \prod_{i=1}^3 \psi_i(J_i) \lrb \sum_{\rhob}
  \Frac{ S^{\J_1}_{\lambdab_1,\rhob}
  S^{\J_2}_{\lambdab_2,\rhob} S^{\J_3}_{\lambdab_3,\rhob}}
  {\bar S_{\vacb,\rhob}}  \,,  \ee
where $\NN$ is the number of triples $(\J_1{,}\J_2{,}\J_3)\iN
\cals_{\lambdab_1}{\times}\cals_{\lambdab_2}{\times}\cals_{\lambdab_3}$ 
such that $\J_1 \J_2 \J_3\eq1$, and where
the matrices $\SJ$ are those introduced in the previous section.

Next we define a matrix $\tC$ with entries
  \be  \tC_{(\lambdab_1,\psi_{\lambda_1}),(\lambdab_2,\psi_{\lambda_2})}
  := \tNl{(\lambdab_1,\psi_1)}{(\lambdab_2,\psi_2)}{\vacb} \,.  \ee
One can show that, up to a normalization, this matrix is a conjugation;
concretely,
  \be  \tC_{(\lambdab,\psi_{\lambda}),(\mub,\psi_{\mu})}
  = \Frac{|\Gs|}{|\cals_\lambda|} \,
  C^{(\lambdab)}_{\psi_\lambda,\psi_\mu}\, \delta_{\lambdab,\mub^+}^{}
  \,, \ee
where $C^{(\lambdab)}_{\psi_\lambda,\psi_\mu}$ is the conjugation on
resolved
fixed points that has been defined in \cite{fusS6}. In particular, $\tC$ is
invertible; we define the structure constants of the classifying algebra
by using the inverse $\tC^{-1}$ as a metric to raise the third index.

In the $\cala$-theory only the fields in the untwisted sector of the 
orbifold theory appear; in terms of the torus, one only has the
twisting by $1\iN\Gs$ in the `space' direction, but projections in 
the `time' direction. A modular S-transformation exchanges `space' and 
`time', thus yielding also the twist sectors; they come without insertion 
in time direction, so the fields are not projected and we have to
consider orbits rather than individual fields. Accordingly,
an important tool in the investigation of the classifying algebra 
$\calc(\calap)$ is a matrix $\tS$ whose row index takes 
values in the set of basis elements of $\calc(\calap)$, while the set for
the column indices consists of pairs $(\rhoB{,}\psi_\rho)$, where $\rhoB$ is 
any $\Gs$-orbit of primary fields of the $\calap$-theory and $\psi_\rho$ is
a character of the (untwisted) stabilizer $\cals_\rho$ of that orbit. 
This matrix $\tS$ takes over the role that the modular 
matrix $S$ of the $\cala$-theory plays for the $\cala$-preserving \bc s. We 
emphasize that {\em all\/} orbits of the $\calap$-theory appear, not just 
the ones with vanishing monodromy charge. Explicitly, $\tS$ is given by
an expression similar to \erf6,
  \be  \tS_{(\lambdab,\psi_\lambda),(\rhoB,\psi_\rho)}
  = \Frac{|\Gs|}{|\cals_\lambda|\,|\cals_\rho|}
  \sum_{\J\in\cals_\rho\cap\cals_\lambda} \psi_\lambda(\J)\,
  \psi_\rho(\J)^*\, \SJ_{\lambdab,\rhob} \,.  \labl Y
One can see that $\tS$ is invertible; the inverse is the matrix with entries
  \be  (\tS^{-1})^{(\rhoB,\psi_\rho),(\lambdab,\psi_\lambda)}_{}
  = \Frac{|\cals_\lambda|}{|\Gs|}\, \tS^*_{(\lambdab,\psi_\lambda),
  (\rhoB,\psi_\rho)} \,.  \ee
In particular $\tS$ is a square matrix, which implies the sum rule 
  \be  \sumbo\lambda |\cals_\lambda| = \sum_\rhoB |\cals_\rho| \,.  \ee
In words, the number of untwisted fields of the $\calap$-theory 
equals the number of all $\Gs$-orbits of fields when both are counted with
multiplicities  given by the number of elements in the stabilizer.

Combining the previous formul\ae\ one checks that the matrix $\tS$ diagonalizes
the matrices of structure constants of the classifying algebra. Put differently,
the structure constants of $\calc(\calap)$ with three lower indices
obey the Verlinde-like formula 
\be  \tNl{(\lambdab_1,\psi_{\lambda_1})}{(\lambdab_2,\psi_{\lambda_2})} 
{(\lambdab_3,\psi_{\lambda_3})} = \sum_\rhoB\sum_{\psi_\rho\in\cals_\rho^*} 
\Frac{ \tS_{(\lambdab_1,\psi_{\lambda_1}),(\rhoB,\psi_\rho)}
  \tS_{(\lambdab_2,\psi_{\lambda_2}),(\rhoB,\psi_\rho)}
  \tS_{(\lambdab_3,\psi_{\lambda_3}),(\rhoB,\psi_\rho)}}
  {\tS_{\vacb,(\rhoB,\psi_\rho)}} \,.  \ee
Also, the conjugation $\tC$ can be expressed through $\tS$ as
  \be \tC_{(\lambdab_1,\psi_1),(\lambdab_2,\psi_2)}^{} =
  \sum_{\rhoB,\psi_\rho} \tS_{(\lambdab_1,\psi_1),(\rhoB,\psi_\rho)}^{}\,
  \tS_{(\lambdab_2,\psi_2),(\rhoB,\psi_\rho)}^{} \,,  \ee
so that after raising the third index we have
  \be  \tN{(\lambdab_1,\psi_1)}{(\lambdab_2,\psi_2)}{(\lambdab_3,\psi_3)}
  = \sum_{\rhoB,\psi_\rho} \Frac{|\cals_{\lambda_3}|}{|\Gs|}\, \Frac{
  \tS_{(\lambdab_1,\psi_1),(\rhoB,\psi_\rho)}\,
  \tS_{(\lambdab_2,\psi_2),(\rhoB,\psi_\rho)}\,
  \tS_{(\lambdab_3,\psi_3),(\rhoB,\psi_\rho)}^* }
  {\tS_{\vacb,(\rhoB,\psi_\rho)}} \,.  \Labl4n
Together these results imply that the classifying algebra $\calc(\calap)$
is commutative and associative, and that the vacuum $\vacb$ is a unit element. 
Since $\calc(\calap)$ is also endowed with a conjugation $\tC$ which is a 
(weighted) evaluation on the identity, it is semi-simple. It follows in
particular that all \irr $\calc(\calap)$-\rep s are \onedim. They are labelled by 
the pairs $(\rhoB,\psi_\rho)$ and
can be neatly expressed in terms of the matrix $\tS$:
  \be  R_{(\rhoB,\psi_\rho)} (\phi_{(\lambdab,\psi_\lambda)}) =
  \frac{\tS_{(\lambdab,\psi_\lambda),(\rhoB,\psi_\rho)}}
  {\tS_{\vacb,(\rhoB,\psi_\rho)}} \,.  \labl R
To summarize: The conformally invariant boundary conditions preserving $\calap$
are in one-to-one correspondence with the pairs $(\rhoB,\psi_\rho)$, and the 
reflection coefficients for any \bc\ are expressible in terms of the matrix 
$\tS$ as in \erf R.

\sect{Automorphism types}
As already mentioned, the monodromy charges \erf4 are
constant on $\Gs$-orbits. This allows us to associate to every $\Gs$-orbit
$\lambdaB$ of the $\calap$-theory a function 
$\calq_\lambdaB{:}\; \Gs{\to}\,\complex$ given by 
  \be  \calq_\lambdaB(\J) = \exp(2\pi\ii Q_\J(\lambdab))  \,.  \ee
The functions $\calq_\lambdaB$ are actually characters on $\Gs$,
i.e.\ elements of the character group $(\Gs)^*_{}$ which can be naturally
identified with the orbifold group, $\calq_\lambdaB\iN{(\Gs)}^*_{}\eq G$.
Thus we can associate to every boundary condition 
$(\rhoB,\psi_\rho)$ an element $\calq_\rhoB$ of the orbifold 
group. We now show that this group element constitutes the {\em automorphism
type\/} \cite{fuSc6,reSC} of the boundary condition. This follows from the fact 
that for every $\J\iN\Gs$ we have
  \be  R_{(\rhoB,\psi_\rho)} (\phi_{(\J\lambdab,\psi_\lambda)}) =
  \Frac{\tS_{(\J\lambdab,\psi_\lambda),(\rhoB,\psi_\rho)}}
  {\tS_{\vacb,(\rhoB,\psi_\rho)}} = \calq_{\rhoB\!}(\J) \,
  \Frac{\tS_{(\lambdab,\psi_\lambda),(\rhoB,\psi_\rho)}}
  {\tS_{\vacb,(\rhoB,\psi_\rho)}} = \calq_{\rhoB\!}(\J) \,
  R_{(\rhoB,\psi_\rho)} (\phi_{(\lambdab,\psi_\lambda)}) \, . \labl3
In particular, the boundary blocks for fields on full orbits contribute to the 
boundary states with a relative phase $\calq_\rhoB(\J)$, 
given by the value of the character $\J\iN\Gs$ on the
group element $\calq_\rhoB$, which can be expressed
by saying that the reflection of a bulk field at the boundary is twisted
by the action of the group element $\calq_\rhoB\iN G$. It follows that indeed 
to any boundary condition one can associate an automorphism of $\cala$, namely 
the one which multiplies the subspace $\bar\calh_{\J\vacb}\, {\subseteq}\,
\calh_\vac$ by $\calq_\rhoB(\J)$.  We stress that this statement
arises as a {\em result} of our analysis rather than being an ad hoc input. 

Also note that 
twisted boundary conditions in the $\cala$-theory are in a 
natural correspondence with the twist sectors of the orbifold theory $\calap$.
(In other words, boundary operators which change the \auto\ type correspond
to the twist fields of the orbifold.) By taking appropriate ideals of 
$\calc(\calap)$,
one can associate an individual classifying algebra $\calc_\calq(\calap)$
to each automorphism type $\calq\iN G$. In particular, for the trivial
\auto\ type $1\iN G$ one recovers the fusion \alg\ of $\cala$.
Individual classifying algebras for non-trivial \auto\ types
where discussed in \cite{fuSc6}; they were
used in \cite{fuSc9} to classify all boundary
conditions of the critical three-state Potts model and to discuss the 
\bc s for other minimal and WZW models with extension modular invariants.
In the special case of the D$_{\rm even}$ (i.e., $\zet_2$-extension) type 
\sltwo\ \wzwts\ at level $k\iN4\zet$, there is a simple closed formula for 
the matrix $\tS$, and the classifying algebra for non-trivial \auto\ type 
can be shown to be isomorphic to the fusion \alg\ of the 
$(\frac k2{+}1\,,2)$ non-unitary Virasoro minimal models.
Incidentally, in this particular case we can also show that the total 
classifying algebra $\calc(\calap)$ is isomorphic to the Pasquier
\cite{pasq4,dizu,pezu} \alg, even though the natural basis for $\calc(\calap)$ 
arising here differs from the one used in Pasquier's context (where 
the diagonalising matrix is taken to be unitary).
It follows in particular that, as also advocated in
\cite{bePz,bppz}, the conformally invariant \bc s\,%
 \futnote{To be precise, at least those which do not correspond to complex 
Chan\hy Paton charges, compare \cite{sasT2}.} 
of the unitary Virasoro minimal models are controlled by the representation
theory of a semi-simple classifying algebra. 

\sect{Annulus coefficients}
Now that we know the reflection coefficients, we would like to compute the
annulus amplitudes. They are linear combinations of characters. One can
show that for an annulus with boundary conditions $(\rhoBe,\psi_1^{})$
and $(\rhoBz,\psi_2^{})$, the characters that appear are those of an integer
spin simple current extension of the $\calap$-theory by the subgroup
  \be  H' \equiv H'_{\!\rho_1\rho_2} := \{\J\iN\Gs \,|\, Q_\J(\rho_1)\eq0\eq
  Q_\J(\rho_2) \} \ee
of $\Gs$. The characters of the primary fields in this extension are
  \be  \X'_{(\rhoB',\psi')}:= \sum_{\J\in H'/\cals'_\rho}\chii_{(\J\rhob,\psi')}
  = \Frac1{|\cals'_\rho|}
  \sum_{\J\in H'} \chii_{(\J\rhob,\psi')} \,,  \labl X
where we introduced $\cals'_\rho\,{:=}\,\cals_\rho{\cap}H'$ and where $\psi'
\iN(\cals'_\rho)_{}^*$.

We would like to know the annulus amplitude in the open string channel. To
this end we have to identify the modular matrix that implements the
transformation of the characters \erf X under $\tau\,{\mapsto}\,{-}1/\tau$;
this is the modular matrix $S'$ of the $H'$-extension as constructed in 
\cite{fusS6} (compare also section 3). Afterwards we define 
the annulus coefficients as the multiplicities of the characters
of the $H'$-extension in the annulus amplitude. We find that
  \be  \A_{(\rhoBe,\psi_1^{})\,(\rhoBz,\psi_2^{})}(t) 
  = \sum_{\sigmaB'}\sum_{\psi'_\sigma\in (\cals'_\sigma)^*_{\phantom|}}
  \A_{(\rhoBe,\psi_1^{})\,(\rhoBz,\psi_2^{})}^{(\sigmaB',\psi'_\sigma)}\,
  \X'_{(\sigmaB',\psi'_\sigma)}(\Frac{\ii t}2)  \ee
with
  \be  \hsp{-.9}\bearll
  \A_{(\rhoBe,\psi_1^{})\,(\rhoBz,\psi_2^{})}^{(\sigmaB',\psi'_\sigma)} \!\!\!
  &= \!\dsty\sumbo\lambda\sum_{\psi_\lambda\in\cals_\lambda^*}
  \!\llb \Frac{ \tS_{(\lambdab,\psi_\lambda),(\rhoBe,\psi_1)}}
  {\tS_{\vacb,(\rhoBe,\psi_1)}}\, \tS_{\vacb,(\rhoBe,\psi_1)} \lrb^{\!*}
  {\cdot}\, \llb \Frac{ \tS_{(\lambdab,\psi_\lambda),(\rhoBz,\psi_2)}}
  {\tS_{\vacb,(\rhoBz,\psi_2)}}\, \tS_{\vacb,(\rhoBz,\psi_2)} \lrb \,
  \\{}\\[-1.75em]
  &\hsp{12.7} \cdot\, \Frac{|\cals_\lambda|}{|\Gs|}\,
  \Frac1{S'_{(\lambdaB',\psi'_\lambda),\vac'}} \cdot
  S'_{(\lambdaB',\psi'_\lambda),(\sigmaB',\psi'_\sigma)}\,
  \\{}\\[-.75em]
  &= \!\dsty\sumbo\lambda\sum_{\psi_\lambda\in\cals_\lambda^*}
  \Frac{|\cals_\lambda|}{|\Gs|}\,
  \Frac{ \tS_{(\lambdab,\psi_\lambda),(\rhoBe,\psi_1)}^*
  \tS_{(\lambdab,\psi_\lambda),(\rhoBz,\psi_2)}\,
  S'_{(\lambdaB',\psi'_\lambda),(\sigmaB',\psi'_\sigma)} }
  {S'_{(\lambdaB',\psi'_\lambda),\vac'}} \,.
  \eear  \labl A
Here the two factors on the \rhs\ of the first line
are products of reflection coefficients and normalizations of vacuum boundary
fields, while the factors in the second line come from the normalization of the 
Ishibashi boundary states and from the modular transformation of the
characters $\chii_{(\lambdab,\psi_\lambda)}$, \resp.
$\psi'_\lambda\iN(\cals_\lambda')^*_{}$ is the restriction of the 
$\cals_\lambda$-character $\psi_\lambda$ to $\cals'_\lambda$.

A crucial property of the annulus multiplicities \erf A is that they are 
non-negative integers; this is required in order to have an interpretation of 
the annulus amplitude as a partition function. Indeed, up to a factor one can 
write the numbers \erf A as a sum of fusion rule coefficients ${\rm N}'$ in 
the $H'$-extension,
  \be \A_{(\rhoBe,\psi_1^{})\,(\rhoBz,\psi_2^{})}^{(\sigmaB',\psi'_\sigma)}
  = \Frac{|H''|\,|\cals'_{\rho_1}|\,|\cals'_{\rho_2}|}
  {|H'|\,|\cals_{\rho_1}|\,|\cals_{\rho_2}|} \cdot \sum_{\J\in\Gs/H''} 
  {\rm N}'{}_{\!(\rhoBz',\psi_2'),(\sigmaB',\psi'_\sigma)}
  ^{\ \ \ \ \ \ \ \ \ \J(\rhoBe',\psi_1')}
  \,,  \Labl24
where $H''\,{:=}\,\cals_{\rho_1}{\cdot}\,
\cals_{\rho_2}{\cdot}\,\cals_{\sigma}{\cdot}\,H'_{\!\rho_1\rho_2}$;
the prefactor can be shown to be integral (for details, see \cite{fuSc+}).
When both \bc s preserve the full bulk symmetry, the formula \Erf24 reduces
to the well-known result that for such \bc s the annulus multiplicities
just coincide with fusion rule coefficients of the $\cala$-theory.

It can also be checked that the annulus multiplicities fulfil further
consistency relations of the usual form. These look most transparent if one
works with $\calap$-characters $\chii_{(\sigmab,\psi_\sigma)}$ in place of the
extended characters $\X'_{(\sigmaB',\psi'_\sigma)}$ of equation \erf X. 
It turns out that the
corresponding coefficients $\tA$ in the annulus amplitude depend only on
the $\Gs$-orbit of $\sigmab$ and are given by
  \be  \tA_{(\rhoBe,\psi_1^{})\,(\rhoBz,\psi_2^{})}^{(\sigmaB,\varphi)}
  = \Frac{|\cals'_\sigma|}{|\cals_\sigma|}\,
  \A_{(\rhoBe,\psi_1^{})\,(\rhoBz,\psi_2^{})}^{(\sigmaB',\varphi')} \ee
($\varphi'$ denotes again the restriction of $\varphi$ to
$\cals'_\sigma$). We then find that, first, the coefficients $\tA$ 
furnish a matrix \rep\ of some algebra,
  \be   \sum_\rhoBd\sum_{\psi_3^{}\in\cals_{\!\rhod}^*}
  \tA_{(\rhoBe,\psi_1^{})\,(\rhoBd,\psi_3^{})}^{(\sigmaBe,\varphi_1^{})}\,
  \tA_{(\rhoBd,\psi_3^{})\,(\rhoBz,\psi_2^{})}^{(\sigmaBz,\varphi_2^{})}
  = \dsty\sum_{\sigmaBd} \sum_{\varphi_3^{}\in\cals_{\sigma_3}^*}
  \tM{(\sigmaBe,\varphi_1^{})}{(\sigmaBz,\varphi_2^{})}{(\sigmaBd,\varphi_3^{})}
  \,\tA_{(\rhoBe,\psi_1^{})\,(\rhoBz,\psi_2^{})}^{(\sigmaBd,\varphi_3^{})} \,,
  \Labl MA
with the structure constants of that \alg\ again given by the $\tA$,
  \be  \tM{(\sigmaBe,\varphi_1^{})}{(\sigmaBz,\varphi_2^{})}
  {(\sigmaBd,\varphi_3^{})}
  = \tA_{(\sigmaBeP,\varphi_1^+)\,(\sigmaBz,\varphi_2^{})}
  ^{(\sigmaBdP,\varphi_3^+)} \,.  \Labl tM
(In particular, according to \Erf24 up to a prefactor the structure constants 
$\TM$ are nothing but sums of fusion rule
coefficients of the $H'$-extension of the $\calap$-theory.)
Note that the \alg\ with structure constants $\TM$ involves orbits 
$\sigmaB$ of arbitrary monodromy charge; the monodromy charge actually
provides a grading of the \alg, with the grade-zero sub\alg\ being 
just the fusion \alg\ of the $\cala$-theory.

Second, the annulus coefficients are `associative' in the sense that\,%
 \futnote{Concerning the use of lower and upper labels for the annulus 
coefficients we stick to the usual convention, compare
e.g.\ \cite{prss2,prss3,fuSc6,reSC}. As indicated by the presence of 
the conjugation on the boundary conditions 
in the formula \Erf AA and in similar relations,
this convention is quite unfortunate.}
  \be  \sum_\sigmaB\sum_{\varphi\in\cals_\sigma^*}
  \tA_{(\rhoBe,\psi_1^{})\,(\rhoBz,\psi_2^{})}^{(\sigmaB,\varphi)}\,
  \tA_{(\rhoBd,\psi_3^{})\,(\rhoBv,\psi_4^{})}^{(\sigmaBP,\varphi^+)}
  = \sum_\sigmaB\sum_{\varphi\in\cals_\sigma^*}
  \tA_{(\rhoBe,\psi_1^{})\,(\rhoBdP,\psi_3^+)}^{(\sigmaB,\varphi)}\,
  \tA_{(\rhoBzP,\psi_2^+)\,(\rhoBv,\psi_4^{})}^{(\sigmaBP,\varphi^+)}
  \,,  \Labl AA
In view of \Erf tM, the two identities \Erf AA and \Erf MA are merely different
manifestations of one and the same relationship.

Finally we mention that, as seen by
comparing the result \erf A with the formula \Erf4n for the structure
constants $\TN$, up to a factor the annulus coefficients are the
`opposite structure constants' for $\calc(\calap)$, i.e.\ those obtained 
when summing over the other index of the non-symmetric diagonalizing
matrix $\tS$.

\sect{Outlook}
To conclude this letter we summarize the structure we found and then
speculate about possible generalizations of this structure. We have seen that
if we require boundary conditions to preserve only the symmetries in an abelian
orbifold subalgebra of the chiral algebra, then the boundary conditions can
be obtained with the help of a natural classifying algebra. 
Moreover, using structures in the corresponding orbifold theory, we could derive
rather than assume that each boundary condition comes with a specific 
automorphism type. The Chan-Paton types \cite{fuSc6} for a given automorphism 
type correspond to simple current orbits in the relevant twist sector
of the orbifold theory. 

It is reasonable to expect that these features will
persist for orbifold subalgebras under a non-abelian group $G$.
For a general consistent subalgebra $\calap$ of $\cala$ which is not given as 
an orbifold subalgebra, we expect that a classifying algebra can be determined 
once the following two pieces of information are available:\\
-- The decomposition of $\cala$-modules in terms of \irr $\calap$-modules.\\
-- An expression of the \cblock s of the $\cala$-theory in terms
   of linear combinations of quotient sheaves of the sheaves of \cblock s
   of the $\calap$-theory.

Another insight is that for any inclusion $\calap\hookrightarrow\cala$
of preserved bulk symmetry algebras, we have a projection of
the corresponding classifying algebras: the classifying algebra for $\cala$
is a quotient of the one for $\calap$. Thus the following picture emerges:
the set $\calm$ of all consistent subalgebras of a given chiral algebra $\cala$
is partially ordered by inclusion. It is reasonable to expect that it is even
an inductive system, i.e.\ given any two consistent subalgebras $\calap_1$ and 
$\calap_2$, one can find a consistent subalgebra $\calap_3$ that is contained
in their intersection, $\calap_3\,{\subseteq}\,\calap_1{\cap}\calap_2$. 
Assuming that also in general for $\calap_1\,{\subset}\,\calap_2$ the 
classifying algebra for $\calap_2$ is a quotient of the one for $\calap_1$, we 
will obtain a projective system of classifying algebras. 
Taking the {\em projective limit\/} over this system,
we obtain a {\em universal classifying algebra\/} which gives all conformally
invariant boundary conditions. This universal classifying algebra can be
explicitly displayed in simple cases, e.g.\ for the free boson compactified 
on a circle or for the $\zet_2$-orbifold of these theories. We are planning 
to come back to a detailed study of this algebra in the future.

\bigskip
\bigskip

\noindent
{\bf Acknowledgement:} We benefited from discussions with the participants of 
the workshop ``Conformal field theory of D-branes" (see {\small\tt
http:/$\!$/www.desy.de/\raisebox{-.2em}{$\tilde{\phantom.}$}jfuchs/CftD.html}).
We are grateful to Bert Schellekens for reading the manuscript.
We also acknowledge the hospitality of the
Erwin-Schr\"odinger Institute in Vienna where part of this work was done. 

\newpage \small
 \newcommand\wb{\,\linebreak[0]} \def\wB {$\,$\wb}
 \newcommand\Bi[1]    {\bibitem{#1}}
 \renewcommand\J[5]     { {\sl #5}, {#1} {#2} ({#3}) {#4} }
 \newcommand\Prep[2]  {{\sl #2}, preprint {#1}}
 \def\jf    {J.\ Fuchs}
 \def\comp  {Com\-mun.\wb Math.\wb Phys.}
 \def\foph  {Fortschr.\wb Phys.}
 \def\ijmp  {Int.\wb J.\wb Mod.\wb Phys.\ A}
 \def\jams  {J.\wb Amer.\wb Math.\wb Soc.}
 \def\joal  {J.\wB Al\-ge\-bra}
 \def\nuci  {Nuovo\wB Cim.}
 \def\nupb  {Nucl.\wb Phys.\ B}
 \def\phlb  {Phys.\wb Lett.\ B}
 \def\phrl  {Phys.\wb Rev.\wb Lett.}
 \def\Bc      {Boundary condition }
 \def\con     {conformal\ }
 \def\dim     {dimension}
 \def\furu    {fusion rule}
 \def\kma     {Kac\hy Moody algebra}
 \def\modinv  {modular invarian}
 \def\Modinv  {Modular invarian}
 \def\parfu   {partition function}
 \def\sym     {symmetry}
 \def\syms    {sym\-me\-tries}
 \def\voa     {vertex operator algebra}
 
 \end{document}